\documentclass{mem}
\usepackage{natbib}\usepackage{txfonts}\usepackage{balance}
\usepackage{graphicx}
\usepackage{amssymb}
\usepackage[a4paper]{hyperref}
\idline{75}{282}
\begin{document}
\def\teff{$T\rm_{eff }$}
\def\kms{$\mathrm {km s}^{-1}$}

\title{Spectropolarimetric investigations of the magnetization of the quiet Sun chromosphere
}

   \subtitle{}

\author{
J. \,Trujillo Bueno\inst{1,2,3}
          }

\offprints{J. Trujillo Bueno}

\institute{
Instituto de Astrof\'{\i}sica de Canarias, 38205, La Laguna, Tenerife, Spain
\and
Departamento de Astrof\'{\i}sica, Universidad de La Laguna, Tenerife, Spain
\and
Consejo Superior de Investigaciones Cient\'{\i}ficas, Spain\\
\email{jtb@iac.es}
}

\authorrunning{Trujillo Bueno}

\titlerunning{The Magnetism of the Quiet Solar Chromosphere}

\abstract{This paper reviews some recent advances in the development and application of polarized radiation diagnostics to infer the mean magnetization of the quiet solar atmosphere, from the near equilibrium photosphere to the highly non-equilibrium upper chromosphere. In particular, I show that interpretations of the scattering polarization observed in some spectral lines suggest that while the magnetization of the photosphere and upper chromosphere is very significant, the lower chromosphere seems to be weakly magnetized. 
}
\maketitle{}

\section{Introduction} 

The chromosphere is a crucial boundary region in the solar outer atmosphere, not only because it is probably the region where the dominant physics changes from hydrodynamic to magnetic forces and most of the non-radiative heating that sustains the corona and solar wind is released, but also because the dissipation of magnetic energy in the $10^6$ K corona may be significantly modulated by the 
strength and structure of the magnetic field in the chromosphere \citep[e.g.,][]{trujillobueno-parker07}.
Unfortunately, our empirical knowledge of the magnetism of the solar outer atmosphere is practically 
non-existent notwithstanding the precious qualitative information provided by high resolution images of the solar atmosphere taken around the wavelengths of strong spectral lines like H$\alpha$ and Ca {\sc ii} 8542 \AA\ \citep[e.g., the review by][]{trujillobueno-rutten07}. Such high cadence, high angular resolution {\em intensity} images demonstrate that the solar chromospheric plasma is extremely inhomogeneous and dynamic and suggest that the upper solar chromosphere is a ``fibrilar dominated-magnetism medium''. They are also useful in helping to constrain the magnetic field orientation, but they do not provide quantitative information on the magnetic field vector because the Stokes $I(\lambda)$ profiles of such strong lines are practically insensitive to its strength, inclination and azimuth. Most probably the magnetic field is the underlying structuring agent, but the fine structure that we see in such intensity images (i.e., the fibrils) directly implies only the presence of thermal and/or density inhomogeneities. 

The only way to obtain quantitative empirical information on the magnetic fields of the extended solar atmosphere is via the measurement and interpretation of the emergent spectral line polarization \citep[e.g.,][] {trujillobueno-stenflobook,trujillobueno-toro,trujillobueno-landilandolfibook}. Solar magnetic fields leave their fingerprints on the polarization signatures of the emergent spectral line radiation. This occurs through a variety of unfamiliar physical mechanisms, not only via the Zeeman effect. In particular, magnetic fields modify the atomic level polarization (population imbalances and quantum coherences) that pumping processes by anisotropic radiation induce in the atoms of the solar atmosphere \citep[e.g.,][]{trujillobueno-jtb01}. Interestingly, this so-called Hanle effect allows us to ``see" magnetic fields to which the Zeeman effect is blind within the limitations of the available and foreseeable instrumentation. We may thus define ``the Sun's hidden magnetism'' as all the magnetic fields of the extended solar atmosphere that are impossible to diagnose via the consideration of the Zeeman effect alone.

A recent review of observational properties of the solar chromosphere was presented by \cite{trujillobueno-judge06}. There are also reviews where the reader finds information on how spectropolarimetric observations allow us to explore chromospheric magnetic fields in quiet and active regions \citep[e.g.,][]{trujillobueno-harvey06,trujillobueno-harvey09,trujillobueno-stenflo06,trujillobueno-lagg07,trujillobueno-casinilandi,trujillobueno-lopezaulanier,trujillobueno-jtb10}. In \cite{trujillobueno-jtb10} I dicuss recent advances in chromospheric and coronal polarization diagnostics, with emphasis 
on the magnetic field of plasma structures embedded in the solar outer atmosphere
(e.g., spicules, prominences, active region filaments and coronal loops). Of particular interest in this respect is the very recent paper by \cite{trujillobueno-centeno10} showing the detection of magnetic fields as strong as 50 G in off-limb spicules of the quiet Sun chromosphere, which could represent a possible lower value of the field strength of organized network spicules at a height of about 2000 km above the visible solar surface.  

In the present paper I focus instead on the diagnostic problem of the  
magnetization of the atmosphere of the ``quiet'' Sun, with emphasis on the variation with height of the 
mean field strength in the quiet chromospheric plasma itself. 
It is important to note that determining the mean magnetization of the quiet Sun requires finding how much flux resides at small scales. To this end, it is crucial to measure and interpret the linear polarization produced by atomic level polarization and its modification by the Hanle effect (see \S2 and \S3). Although in the quiet Sun the amplitudes of such linear polarization signals are often larger than those of the $V/I$ profiles produced by the longitudinal Zeeman effect, their measurement with the available telescopes still requires to sacrifice the spatio-temporal resolution to be able to reach the required polarimetric sensitivity. For this reason, with present telescope apertures, the first step is to try to obtain information on the mean intensity, $\langle B \rangle$, of the actual distribution of magnetic field strengths. The shape of the ensuing probability distribution funtion, PDF($B$), describing the fraction of quiet Sun plasma occupied by magnetic fields of strength $B$, is difficult to determine empirically, although numerical experiments of magnetoconvection suggest that assuming an exponential shape for the PDF is a suitable approximation that avoids overestimating $\langle B \rangle$. Nevertheless, here I consider mainly the simplest model of a single value field that fills the entire atmospheric volume (i.e., PDF($B$)=${\delta}(B-\langle B \rangle)$), with the aim of drawing at least some preliminary conclusions on the lower limit for $\langle B \rangle$ in the photosphere (\S4), upper chromosphere (\S5) and lower chromosphere (\S6). 
In terms of the heights $h$ (in km) above the visible solar surface where the spectral lines used here are sensitive to the local atmospheric conditions we have $200\,{\lesssim}\,h\,{\lesssim}\,400$ for the photosphere, $1800\,{\lesssim}\,h\,{\lesssim}\,2200$ for the ``upper chromosphere'', and $900\,{\lesssim}\,h\,{\lesssim}\,1300$ for the ``lower chromosphere''.

\section{Pros and Cons of the Hanle and Zeeman Effects}

The polarization of the Zeeman effect is due to the wavelength shifts between the $\pi$ and $\sigma$ transitions composing a spectral line. The typical observational signature of the circular polarization produced by the longitudinal Zeeman effect is an antisymmetric Stokes $V(\lambda)$ profile whose amplitude scales with the ratio, ${\cal R}$, between the Zeeman splitting and the Doppler broadened line width. The linear polarization amplitudes of the transverse Zeeman effect scale instead as ${\cal R}^2$ and its characteristic  observational signatures are symmetric Stokes $Q(\lambda)$ and $U(\lambda)$ profiles with their wing lobes of opposite sign to the line center one. Due to cancellation effects the polarization of the Zeeman effect as a diagnostic tool tends to be blind to magnetic fields that are randomly oriented on scales too small to be resolved. Note also that 

\begin{equation}
{\cal R}={{1.4{\times}10^{-7}{\lambda}B}\over{\sqrt{1.663{\times}10^{-2}T/\alpha+{\xi}^2}}},
\end{equation}
where $\alpha$ is the atomic weight of the atom under consideration and
$\lambda$ is in \AA, $B$ in gauss, $T$ in K and the microturbulent velocity $\xi$ in kms$^{-1}$ \citep[see][]{trujillobueno-landilandolfibook}.

In the quiet solar atmosphere ${\cal R}{\ll}1$ (e.g., ${\cal R}{\approx}10^{-2}$ for H$\alpha$ and ${\cal R}{\approx}5{\times}10^{-2}$ for the Ca {\sc ii} 8542 \AA\ line), which explains why it is far more difficult to detect the signature of the transverse than the longitudinal Zeeman effect in strong chromospheric lines. In practice, only the impact of the Zeeman effect on Stokes $V$ is detected, and mainly in near-IR lines like the 8542 \AA\ line of Ca {\sc ii}. However, the response function of the emergent Stokes $V$ to magnetic field perturbations at various heights in models of the quiet solar atmosphere indicates that the circular polarization produced by the Zeeman effect in spectral lines like H${\alpha}$ and Ca {\sc ii} 8542 \AA\ is insensitive to the physical conditions of the upper chromosphere \citep[e.g.,][]{trujillobueno-snjtbbrc,
trujillobueno-socasuitenbroek04,trujillobueno-uitenbroek06}. For example, Fig. 1 illustrates that in the quiet Sun the circular polarization of the H$\alpha$ line is sensitive mainly to the photospheric magnetic field  \citep[see also][]{trujillobueno-socasuitenbroek04}. The emergent Stokes $V$ profiles in the k-line of Mg {\sc ii} and in Ly${\alpha}$ show a more favourable sensitivity to magnetic fields in the upper solar chromosphere and transition region, but the expected Stokes $V$ signals are very weak (see Eq. 1). In summary, the Zeeman effect is of limited practical interest for the exploration of magnetic fields in the solar outer atmosphere (chromosphere, transition region and corona).

\begin{figure}[t!]
\resizebox{\hsize}{!}{\includegraphics[width=1.0\textwidth]{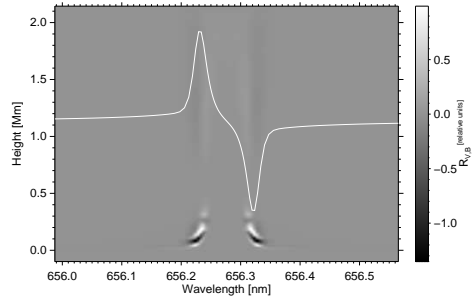}}
\caption{\footnotesize
Stokes $V$ response function of the H$\alpha$ line to  
magnetic strength perturbations in the FAL-C atmospheric model of \cite{trujillobueno-falc}, assuming that it is permeated by a 1000 G vertical field.  
Courtesy of H. Uitenbroek.}
\label{fig:fig-4}
\end{figure}  

Fortunately, there is yet another physical mechanism by means of which 
the magnetic fields of the solar atmosphere leave fingerprints 
on the polarization of the emergent spectral line radiation: the Hanle effect.
Anisotropic radiation pumping processes produce atomic level polarization (i.e., population imbalances and quantum coherences among the magnetic sublevels pertaining to any given degenerate energy level). The Hanle effect can be defined as any modification of the atomic level polarization due to the presence of a magnetic field, including the remarkable effects produced by the level crossings and repulsions that take place when going from the Zeeman effect regime to the complete Paschen-Back effect regime \citep[e.g.,][]{trujillobueno-belluzzi07}. The Hanle effect is especially sensitive to magnetic strengths between $0.1\,B_H$ and $10\,B_H$, where 

\begin{equation}
B_{\rm H}=(1.137\times10^{-7})/(t_{\rm life}\,g_J)
\end{equation}
is the critical Hanle field intensity (in gauss) for which the Zeeman splitting of the $J$-level under consideration is similar to its natural width. Note that $g_J$ 
is the level's Land\'e factor and $t_{\rm life}$ its radiative lifetime in seconds.
Since the lifetimes of the upper levels of the transitions of interest are usually much smaller than those of the lower levels, clearly diagnostic
techniques based on the lower-level Hanle effect are sensitive to much
weaker fields than those based on the upper-level Hanle effect.

The main properties of the Hanle effect are:

{\bf (a)} The Hanle effect is sensitive to weaker magnetic fields than the Zeeman effect (from at least 1~mG to a few hundred gauss), regardless of how large the line width due to Doppler broadening is. For stronger fields, the Hanle effect remains sensitive to the magnetic field orientation. Moreover, the Hanle effect is sensitive to magnetic fields that are randomly oriented on scales too small to be resolved \citep[][]
{trujillobueno-stenflo82,trujillobueno-jtbnature04}.

{\bf (b)} The Hanle effect as a diagnostic tool is {\em not\/} limited to a narrow solar limb
zone. In particular, in the forward scattering geometry of a solar disk center observation,
the Hanle effect creates linear polarization in the presence of an  
inclined magnetic field \citep[e.g.,][]{trujillobueno-jtbnature02}.

{\bf (c)} The Hanle effect operates in the line core and the ensuing response function of the 
emergent linear polarization to magnetic field perturbations shows that in some spectral lines 
(e.g., Ca {\sc ii} 8542 \AA, H$\alpha$, Mg {\sc ii} k and Ly$\alpha$) is sensitive to the magnetic fields of the upper chromosphere and transition region \citep[see a H$\alpha$ response function in figure 4 of][]{trujillobueno-stepan10a}.

In summary, the Hanle effect in strong spectral lines is the key physical mechanism that should be increasingly exploited for quantitative explorations of the magnetism of the solar chromosphere.

\section{Forward modeling of the spectral line polarization}

In general, the modeling of the Stokes profiles in strong lines like H$\alpha$ and the IR triplet of Ca {\sc ii} requires solving a rather complicated radiative transfer problem, known as the non-LTE problem of the
$2^{\rm nd}$ kind \citep[][]{trujillobueno-landilandolfibook}.
This consists in calculating, at each spatial point of any given atmospheric model and
for each $J$-level of the chosen atomic model, the diagonal and non-diagonal elements of the
atomic density matrix that are consistent with the intensity, 
polarization and symmetry properties of the radiation field generated within the (generally
magnetized) medium under consideration. Once such density matrix
elements are known it is straighforward to solve the Stokes vector transfer
equation to obtain the emergent Stokes profiles for any desired line of sight. Highly efficient iterative methods and accurate formal solvers of the Stokes vector transfer equation were developed for solving this type of (complete redistribution) multilevel radiative transfer problem \citep[see the review by][]{trujillobueno-jtb03}. Such methods have been implemented in multilevel computer programs for the generation and transfer of polarized radiation \citep[][]{trujillobueno-msjtbspw3,trujillobueno-stepan08,trujillobueno-jtbshu07,trujillobueno-jtbshu09}. Moreover, the same radiative transfer methods have been recently generalized by \cite{trujillobueno-sampoorna10} for solving the two-level atom problem of 
resonance line polarization taking into account partial redistribution effects, which may be a suitable approximation for modeling the fractional linear polarization profiles in Ly$\alpha$ and Mg {\sc ii} $k$. 

The following sections discuss how the modeling of spectropolarimetric observations through the application of the above-mentioned radiative transfer codes allows us to obtain information on the mean magnetization of the 
quiet solar atmosphere.

\section{The magnetization of the photosphere of the quiet Sun}

The linear polarization profiles produced by scattering processes in the quiet solar atmosphere 
have been observed with poor spatial and/or temporal resolution \citep[e.g.,][]{trujillobueno-stenflokeller97,trujillobueno-gan00}. For this reason,  
\cite{trujillobueno-jtbnature04} confronted observations of the center-to-limb variation of the scattering polarization in the photospheric Sr {\sc i} 4607 \AA\ line with calculations of the $Q/I$ profiles that result from spatially averaging the emergent $Q$ and $I$ profiles calculated in a three-dimensional (3D) model of the quiet solar photosphere resulting from realistic hydrodynamical simulations of solar surface convection. The very significant discrepancy between the calculated and the observed polarization amplitudes indicated the ubiquitous existence of tangled magnetic fields in the quiet solar photosphere, with a mean strength significantly larger than derived from simplistic one-dimensional radiative transfer investigations \citep[see the review by][]{trujillobueno-jtbspw4}. The inferred mean strength of this hidden field turned out to be $\langle B \rangle {\sim} 100$\,G (see Fig. 2), which implies an amount of magnetic energy density that is more than sufficient to compensate the energy losses of the solar outer atmosphere. This estimation was obtained by using the approximation of a {\em microturbulent} field (i.e., that the hidden field has an isotropic distribution of orientations within a photospheric volume given by ${\cal  L}^{3}$, with ${\cal  L}$ the mean-free-path of the {\em line-center} photons). Calculations based on the assumption that the unresolved magnetic field is instead horizontal also lead to the conclusion of a sizable $\langle B \rangle$ \citep[see \S~4 in][]{trujillobueno-jtbspw4}.

What is the physical origin of this ``hidden'' magnetic field whose reality is now being supported by \cite{trujillobueno-lites08} and \cite{trujillobueno-orozco07} through high-spatial-resolution observations of the Zeeman effect taken with Hinode? Is it mostly the result of dynamo action by near-surface convection, as suggested by \cite{trujillobueno-cattaneo99}? Or is it dominated by small-scale flux emergence from deeper layers and recycling by the granulation flows? The fact that the inferred magnetic energy density is a significant fraction (i.e., ${\sim}20\%$) of the kinetic energy density, and that the scattering polarization observed in the Sr {\sc i} 4607 \AA\ line does not seem to be modulated by the solar cycle, strongly supported the suggestion that a surface dynamo plays a significant role for the quiet Sun magnetism \citep[see][]{trujillobueno-jtbnature04}. Recent radiative MHD simulations of dynamo action by near-surface convection also support this possibility \citep[]{trujillobueno-vogler07}.

\begin{figure}[t!]
\resizebox{\hsize}{!}{\includegraphics[width=0.5\textwidth]{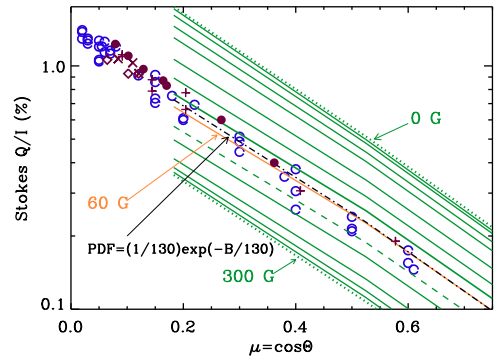}}
\caption{\footnotesize
Center-to-limb variation of the $Q/I$ scattering amplitudes of the photospheric line of Sr {\sc i} at 4607 \AA. Note that $\mu={\rm cos}{\theta}$, with $\theta$ being the heliocentric angle. 
The symbols correspond to various observations taken by several authors during a minimum and a maximum of the solar activity cycle. The dotted, dashed and solid lines (colored in the online version) show scattering polarization calculations in a 3D hydrodynamical model of the photosphere, assuming a volume-filling, single-value microturbulent field with the following magnetic strengths (in G): 0, 5, 10, 15, 20, 30, 40, 50, 60, 80, 100, 150, 200, 250 and 300. The best overall fit to the observations is obtained for 60 G, but note that the best fit is achieved with 70 G at $\mu=0.6$ ($h{\approx}200$ km) and with 50 G at $\mu=0.1$ ($h{\approx}400$ km). With a more realistic exponential probability distribution function (black, dashed-dotted line), the best overall fit is obtained for $\langle B \rangle {=} 130$\,G, and note again that there is a clear indication that $\langle B \rangle$ increases with depth. These results demonstrate that there is a vast amount of ``hidden'' magnetic energy in the quiet solar photosphere, which is much more than sufficient to balance the radiative energy losses of the solar outer atmosphere. From \cite{trujillobueno-jtbnature04}.
}
\label{fig:fig-1}
\end{figure}

In summary, the small-scale magnetic activity of the quiet Sun photosphere is indeed very significant and might be important for understanding the propagation of energy into the outer atmosphere and the flux emergence process. The possibility that with the ``Zeeman eyes of Hinode" we might still be seeing 
only the ``tip of the iceberg'' of the quiet Sun magnetism 
is not surprising because 80\% or more of the vertical unsigned flux seems to be invisible to observations of 
the Zeeman effect at Hinode's resolution of 200 km owing to the cancellation of the 
Stokes $V$ signal from opposite magnetic polarities (Pietarila Graham et al. 2009; see also Stenflo 1994; 
Emonet \& Cattaneo 2001; S\'anchez Almeida et al. 2003; Mart\'\i nez Gonz\'alez et al. 2010).

%
%

\section{The magnetization of the upper chromosphere of the quiet Sun}

There are various spectral lines whose $Q/I$ and $U/I$ profiles are sensitive to the magnetization of the upper chromosphere of the quiet Sun, such as those considered below. It is, however, necessary to emphasize that determining $\langle B \rangle$ from the observed fractional linear polarization signals usually requires confronting them with those that the highly inhomogeneous and dynamic solar chromospheric plasma would produce if it were unmagnetized. This strategy could be applied successfully for determining the mean magnetization of the quiet solar photosphere by solving the radiative transfer problem for the Sr {\sc i} 4607 \AA\ line in a realistic three-dimensional (3D) hydrodynamical model (see \S4), but a similar approach for instead inferring the magnetization of the quiet chromospheric plasma is not yet possible mainly because to produce a realistic 3D model of the thermal, density and dynamic structure of the quiet chromosphere is still computationally prohibitive \citep[e.g.,][]{trujillobueno-carlsson07}. As a matter of fact, the current 3D models do not show fibrils in Ca {\sc ii} 8542 \AA\ and the synthetic H$\alpha$ line-center intensity images show instead the granulation pattern \citep[e.g.,][]{trujillobueno-leenaarts10}. Should we then abandon any attempt to infer the magnetization of the quiet chromosphere via multilevel radiative transfer modeling using the available 1D semi-empirical models ? In my opinion, the solar chromosphere is such an important region that we should at least try to do something potentially useful in spite of the obvious fact that any such 1D model is a poor representation of the complex chromospheric conditions.  

\subsection {The Ca~{\sc ii} 8542 \AA\ line}

The circular polarization of the Ca~{\sc ii} IR triplet is caused by the longitudinal Zeeman effect.
With the available telescopes 
the ensuing $V/I$ signals are measurable even in quiet regions, where their amplitudes are   
${\sim}10^{-3}$ and smaller. Unfortunately, the Zeeman effect in the IR triplet of Ca {\sc ii} is of 
little practical interest for investigating the magnetism of the upper solar chromosphere. Calculations of the Stokes $V$ response function of the strongest line of the Ca {\sc ii} IR triplet to perturbations in the magnetic field strength show that in semi-empirical models of the quiet solar atmosphere the emergent circular polarization is sensitive only to changes between 700 and 1200 km, approximately \citep[e.g.,][]{trujillobueno-uitenbroek06}. 

In quiet regions the linear polarization of the Ca~{\sc ii} IR triplet is dominated by atomic level polarization and its modification by the Hanle effect. Typically, the ensuing $Q/I$ and $U/I$ profiles have their maximum values at the line center. While the linear polarization in the 8498 \AA\ line shows sensitivity to inclined magnetic fields with strengths between 1 mG and 50 G, the emergent linear polarization in the 8542~\AA\ and 8662~\AA\ lines is sensitive to magnetic fields with strengths in the milligauss range (see Fig. 3). The reason for this very interesting behavior is that the scattering polarization in the 8498 \AA\ line gets a significant contribution from the selective emission processes that result from the atomic polarization of the short-lived upper level, while that in the 8542~\AA\ and 8662~\AA\ lines is dominated by the selective absorption processes that result from the atomic polarization of the metastable (long-lived) lower levels \citep{trujillobueno-msjtbprl,trujillobueno-msjtbcoimbra}. Therefore, in quiet regions of the solar atmosphere the magnetic sensitivity of the linear polarization of the 8542~\AA\ and 8662~\AA\ lines is controlled by the lower-level Hanle effect, which implies that in regions with $B>1$ G their Stokes $Q$ and $U$ profiles are only sensitive to the orientation of the magnetic field vector.

\begin{figure*}
\includegraphics[width=\textwidth]{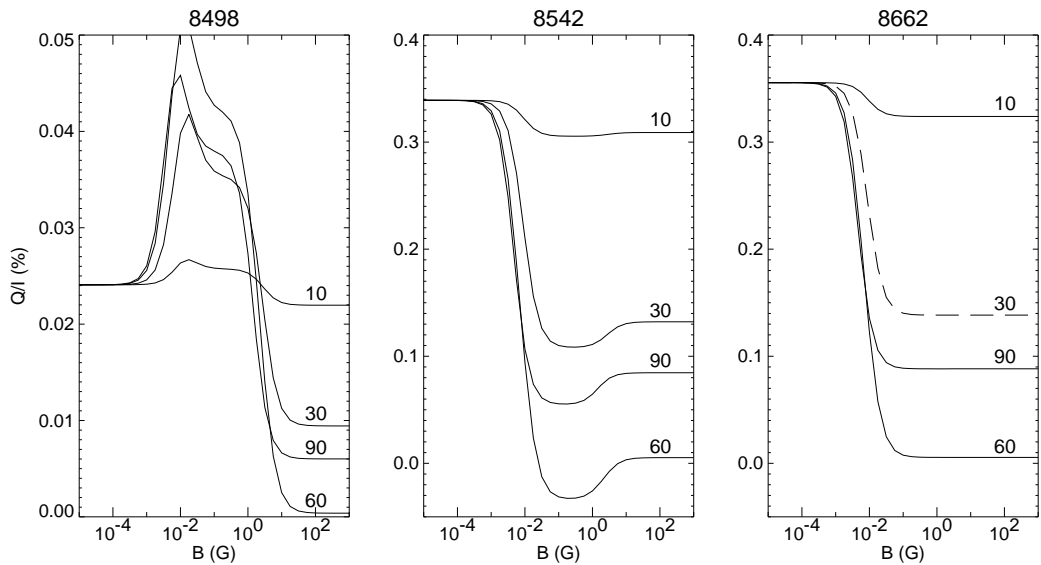}
\caption{\footnotesize
The emergent fractional linear polarization line-center amplitudes  
of the Ca {\sc ii} IR triplet calculated for a line of sight with $\mu=0.1$ in the FAL-C model of the solar atmosphere. Each curve corresponds to the indicated inclination, in degrees with respect to the solar local vertical direction, of the assumed random-azimuth magnetic field. From \cite{trujillobueno-msjtb10}. 
}
\label{fig:fig-2}
\end{figure*}

The most diagnostically interesting lines 
of the Ca {\sc ii} IR triplet are the strongest and the weakest (i.e., the 8542 \AA\ and the 8498 \AA\ lines, respectively). Their linear polarization signals resulting from atomic level
polarization and the Hanle effect can be exploited to
explore the thermal and magnetic structure of the solar
chromosphere. They should also be used to evaluate the degree of realism
of 3D magnetohydrodynamic simulations of the chromosphere
via careful comparisons of the Stokes profiles obtained through forward modeling 
calculations with those observed in quiet regions (e.g., like the ones in Fig. 4). As mentioned above, 
the current 3D models do not show fibrils in the synthetic intensity images calculated at the core of Ca {\sc ii} 8542 \AA, in spite of the fact that the snapshot chosen by \cite{trujillobueno-leenaarts09} 
has $\langle B \rangle {=} 150$\,G  \citep[i.e., a value in agreement with that inferred by][]{trujillobueno-jtbnature04}. 

\begin{figure*}
\includegraphics[width=\textwidth]{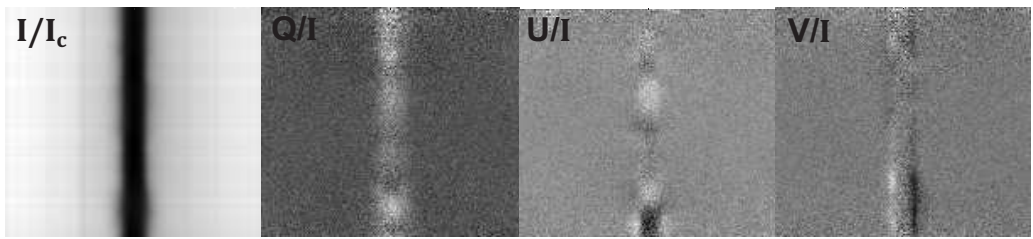}
\caption{\footnotesize
An example of our recent spectropolarimetric observations of the
Ca {\sc ii} 8542 \AA\ line in a very quiet region close to the solar
limb, using ZIMPOL at the Franco-Italian telescope THEMIS. 
The reference direction for Stokes $Q$ is the tangent to the closest limb. 
From \cite{trujillobueno-jtbthemiszimpol}.
}
\label{fig:fig-3} 
\end{figure*}

In particular, the linear polarization of the 8542 \AA\ line should  
be increasingly exploited to explore the magnetic field structure of the
upper chromosphere, ideally via high angular resolution two-dimensional (2D) 
spectropolarimetry with large aperture telescopes 
and novel instruments like IBIS, CRISP or the G\"ottingen Fabry-Perot.  
One drawback is that for $B>1$ G the scattering polarization of the Ca {\sc ii} 8542 \AA\ line
is sensitive only to the orientation of the magnetic field vector. Therefore, in principle, from the spatial variations of the $Q/I$ and $U/I$ signals we can see in Fig. 4 we can only say that they are probably due to changes in the orientation of the magnetic field in the upper chromosphere of the quiet Sun. Although
the spatio-temporal resolution of this spectropolarimetric observation
is rather low (i.e., no better than $3\arcsec$ and 20 minutes), the
fractional polarization amplitudes fluctuate between $10^{-4}$ and $10^{-3}$ 
along the spatial direction of the spectrograph slit, with a typical
spatial scale of $5\arcsec$. Interestingly enough, while the Stokes
$Q/I$ signal changes in amplitude but remains always positive 
along that spatial direction, the sign of the Stokes $U/I$ signal
fluctuates. This is compatible with the action of the Hanle effect in  
a magnetized plasma with $B>1$ G and having 
a spatially varying magnetic field azimuth, which in turn is consistent with the possibility that 
the fibrils seen in the high-resolution intensity images taken by \cite{trujillobueno-cauzzi08} 
at the line center of ${\lambda}8542$ trace out magnetic lines of force.

\subsection {The H${\alpha}$ line}

As we have seen, the linear polarization of the Ca {\sc ii} 8542 \AA\ line is sensitive to the orientation of the 
magnetic field in the upper chromosphere of the quiet Sun, but not to its strength unless $B<1$ G there.
In order to obtain empirical information on the magnetic strength in the upper chromosphere   
we need to use similarly strong lines, but such that their scattering polarization is sensitive to magnetic strengths in the gauss range. Among the various possible choices, H$\alpha$ is particularly interesting because it reaches the Hanle saturation regime for $B{\gtrsim}50$ G and the shape of its fractional scattering polarization profile is very sensitive to the presence of magnetic field gradients in the upper chromosphere of the quiet Sun \citep[][]{trujillobueno-stepan10b}. Moreover, the fibrilar nature of the upper chromosphere is seen even more clearly in H$\alpha$, especially when observing quiet regions far away from the solar disk center \citep[e.g., see figure 7 of][]{trujillobueno-rutten07}. In the remaining part of this section I summarize the main results of this recent paper by \cite{trujillobueno-stepan10b}. 

The temperature minimum region of solar atmospheric models is transparent to H${\alpha}$ radiation \citep[][]{trujillobueno-schoolman72}. As a result, we see the upper chromosphere at the very line center of the H${\alpha}$ line but the photosphere in the line wings. It is thus not surprising what Fig. 1 shows for H$\alpha$, namely that the response of the emergent circular polarization induced by the Zeeman effect
to magnetic field perturbations exhibits large photospheric contributions. Moreover, in the quiet Sun the $V/I$ signals are very weak (${\sim}10^{-4}$ and smaller).

On the contrary, in quiet regions the linear polarization observed in H${\alpha}$ is fully dominated by the presence of radiatively induced population imbalances and quantum coherences among the magnetic sublevels of the line's levels, which produce linear polarization profiles whose maximum values are located at the very line center. The fractional polarization amplitudes vary between $10^{-3}$ and $10^{-4}$. Moreover, this scattering line polarization is modified by the Hanle effect, which operates mainly in the line core and gives rise to $Q/I$ and $U/I$ profiles different from those corresponding to the zero-field case. The response function of the 
emergent linear polarization to magnetic field perturbations shows that the Hanle effect in H${\alpha}$
is sensitive to the magnetic fields of the upper chromosphere \citep[see figure 4 of][]{trujillobueno-stepan10a}.

\begin{figure*}[t!]
\includegraphics[width=0.5\textwidth]{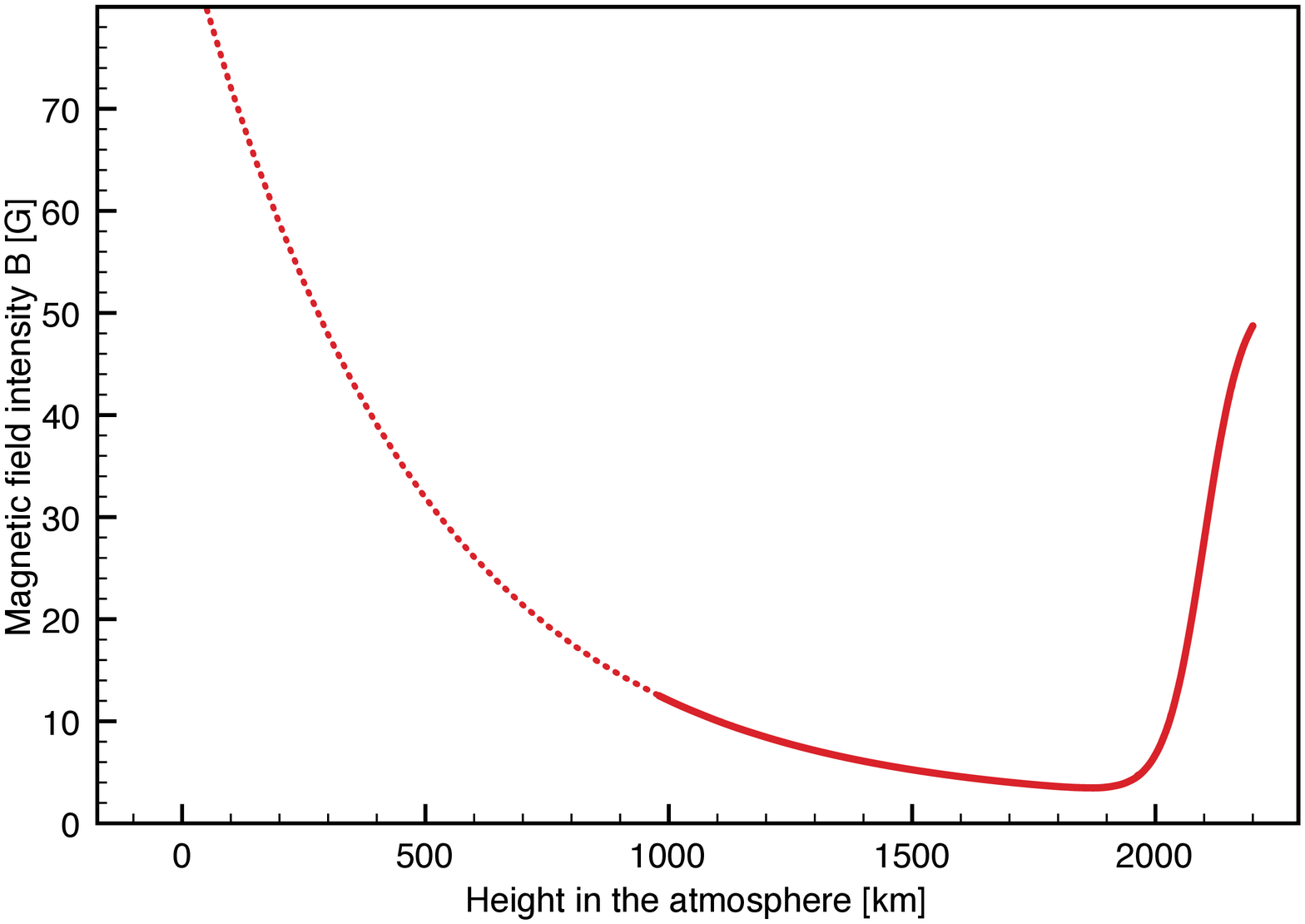}%
\includegraphics[width=0.5\textwidth]{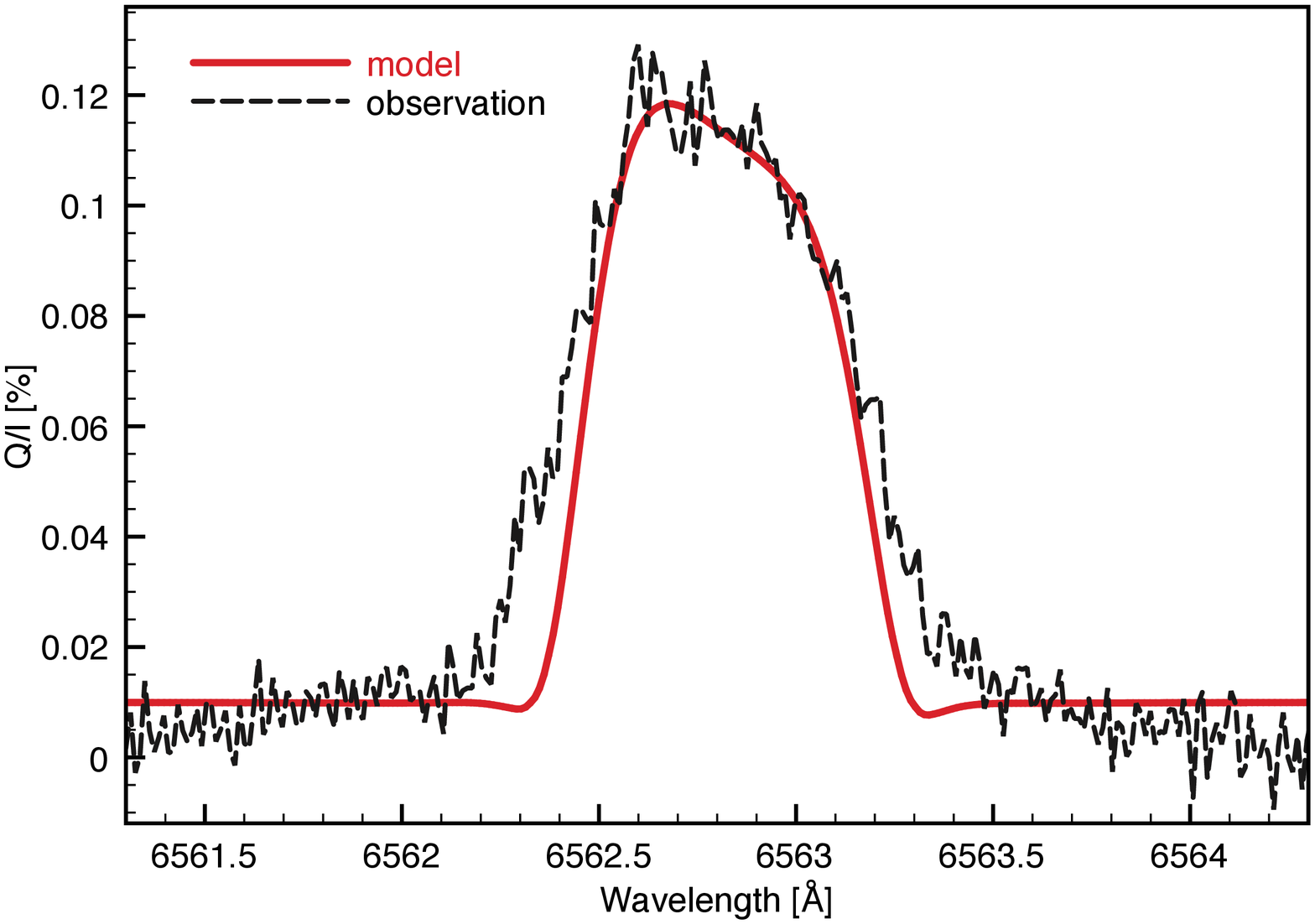}
\caption{\footnotesize
Magnetic field strength model (left panel) and calculated vs. observed $Q/I$ profiles (right panel). 
In the right panel the dashed line shows the $Q/I$ profile observed by \cite{trujillobueno-gan00}, while the
solid line shows the $Q/I$ profile calculated by solving the multilevel scattering polarization problem in the presence of the Hanle effect produced by a significantly inclined magnetic field having at each height a random azimuth and the strength given in the left panel. The total $Q/I$ profiles include the contribution of the continuum polarization. Note that for a line of sight with $\mu=0.1$ 
the scattering polarization of the H${\alpha}$ line is sensitive to the structure of the magnetic field only in the atmospheric region indicated by the solid line part of the model. For more information see \cite{trujillobueno-stepan10b}.
}
\label{fig:fig-5}
\end{figure*}

In spite of its modeling difficulties, 
the Hanle effect in H${\alpha}$ should be exploited for facilitating quantitative explorations of the magnetism of the upper solar chromosphere. A first step has been recently taken by \cite{trujillobueno-stepan10b}. The dashed line in the right panel of Fig. 5 shows the spatially and temporally averaged $Q/I$ profile observed by \cite{trujillobueno-gan00} in a quiet region at about 5 arcseconds from the solar limb. Its more peculiar feature is the asymmetry that it shows around the line center, which cannot be explained by the mere fact that the H$\alpha$ line results from the superposition of seven blended components, four of which making a significant contribution to Stokes $Q$. In their paper \cite{trujillobueno-stepan10b} argue that 
the observed $Q/I$ profile can be explained by the Hanle effect of an inclined magnetic field whose mean strength varies with height as approximately indicated in the left panel of Fig. 5. This suggests the presence of an abrupt and significant magnetization in the upper chromosphere of the quiet Sun and that the average ratio $\beta$ of gas to magnetic pressure decreases suddenly there. 

\section{The magnetization of the lower chromosphere of the quiet Sun}

The magnetic field model in the left panel of Fig. 5 is characterized
by a magnetic complexity zone with $\langle B \rangle {>} 30$\,G in the upper solar chromosphere (i.e., just below the sudden transition region to the $10^6$ K coronal temperatures) and by a strongly 
magnetized photosphere and a weakly magnetized lower chromosphere. 
The suggested abrupt magnetization in the upper chromosphere of the quiet Sun is introduced to 
produce a $Q/I$ profile with a line center asymmetry similar to that found in the observed profile. The strong magnetization of the model's photospheric region is strongly supported by the 3D radiative transfer modeling of the scattering polarization observed in the Sr {\sc i} 4607 \AA\ line \citep[see][]{trujillobueno-jtbnature04}, which indicates that the bulk of the quiet solar photosphere is teeming with a distribution of tangled magnetic fields having a mean strength $\langle B \rangle {\approx} 60$\,G (when estimating $\langle B \rangle$ assuming the simplest case, adopted in \S5 and \S6 of this paper, of a single value field strength). Is the mean magnetization of the lower chromosphere really as weak as indicated in Fig. 5 (i.e., with $\langle B \rangle{\lesssim}10$~G around a height of 1000 km) ?

\begin{figure*}[t!]
\includegraphics[width=0.5\textwidth]{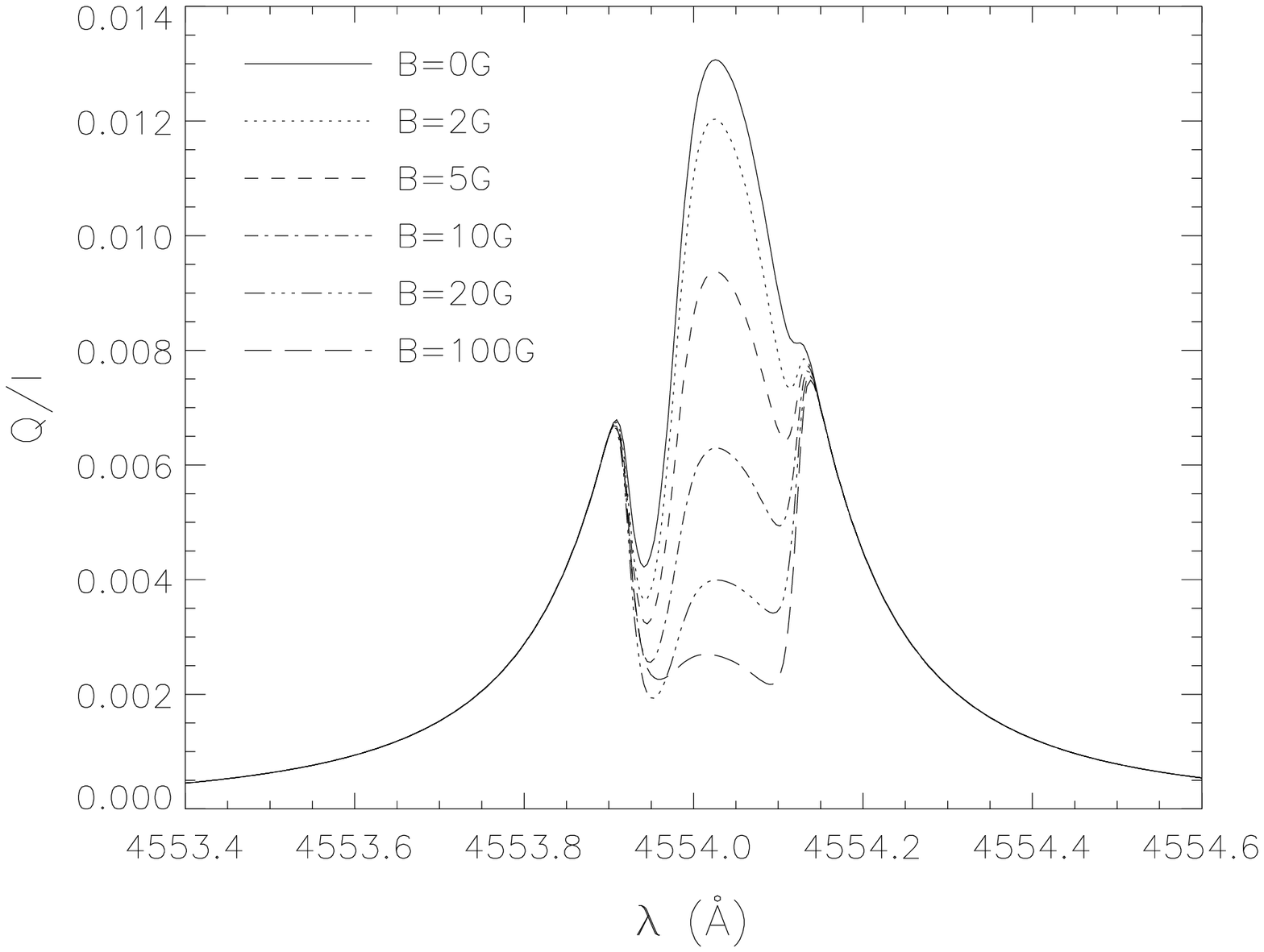}%
\includegraphics[width=0.5\textwidth]{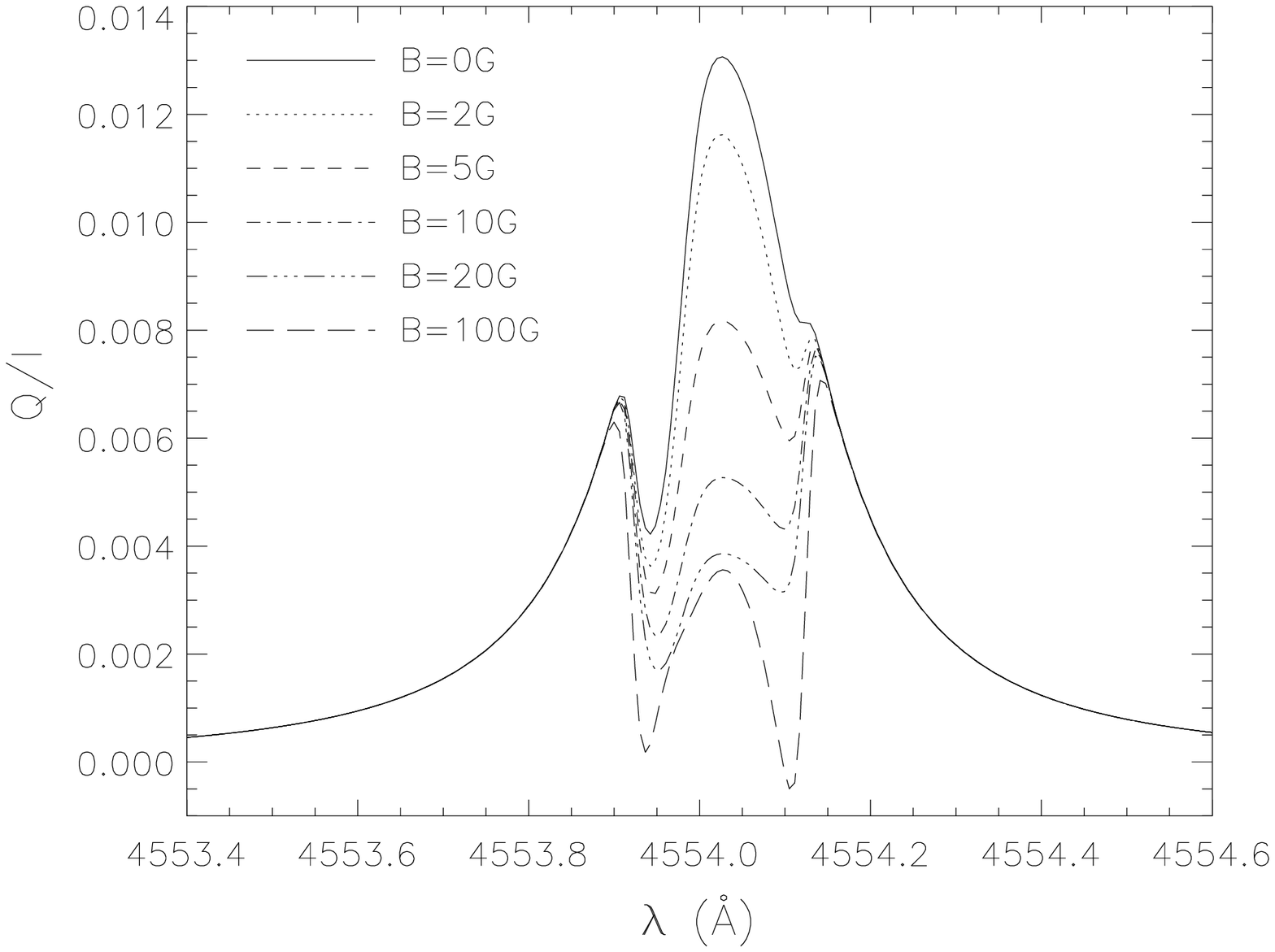}
\caption{\footnotesize
Theoretical estimate of the emergent $Q/I$ profiles of the Ba {\sc ii} D$_2$ line in $90^{\circ}$ scattering geometry, assuming the presence of a microturbulent and isotropic field (left panel) and a horizontal field of random azimuth (right panel). The solid line is very similar to the $Q/I$ profile observed by \cite{trujillobueno-stenflokeller97} in a quiet region close to the solar limb. For more information see \cite{trujillobueno-belluzzi07}.
}
\label{fig:fig-6}
\end{figure*}

One spectral line whose observed scattering polarization supports the possibility of a weakly magnetized lower chromosphere is the D$_2$ line of Ba {\sc ii} at 4554 \AA\ \citep[see][]{trujillobueno-belluzzi07}. Figure 6 shows the sensitivity of the emergent $Q/I$ profile to the magnetic field strength, both for the case of a microturbulent and isotropically distributed magnetic field (left panel) and for the case of a random-azimuth horizontal field (right panel). In fact, the $Q/I$ profile observed by \cite{trujillobueno-stenflokeller97} has a three-peak structure, very similar to that shown by the solid lines in Fig. 6. Note that for $B{\gtrsim}10$ G the amplitude of the central $Q/I$ peak is smaller than the amplitudes of the wing peaks, contrary to what the observed $Q/I$ profile shows. Another spectral line whose linear polarization suggests that the lower chromosphere of the quiet Sun has a small $\langle B \rangle$ value is the Na {\sc i} D$_1$ line \citep[see the review by][]{trujillobueno-jtbspw5}.

\section{Concluding comments}

Three are the main conclusions to be drawn here:

{\bf (1)} The bulk of the quiet solar photosphere is strongly magnetized, with   
$\langle B \rangle{\sim}$100 G when no distinction is made between granules and intergranules.

{\bf (2)} The lower chromosphere seems to be weakly magnetized, with $\langle B \rangle{<}10$~G.

{\bf (3)} The magnetization of the upper chromosphere of the quiet Sun is abrupt and significant, with $\langle B \rangle {>} 30$\,G just below the sudden transition region to the $10^6$ K coronal plasma.

Are these conclusions reliable? They are based on radiative transfer modeling of the 
scattering polarization $Q/I$ profiles of some spectral lines 
observed without spatial and/or temporal resolution in quiet regions 
close to the edge of the solar disk. The radiative transfer calculations have been carried 
out in given atmospheric models (see below). Such $Q/I$ signals depend on the anisotropy of the 
radiation field within the solar atmosphere, which is sensitive to its thermal and density structure.
The $Q/I$ amplitudes are also sensitive to collisions with neutral hydrogen atoms (e.g., the case of the Sr {\sc i} 4607 \AA\ line) and protons (e.g., the case of H$\alpha$). Through the Hanle effect the emergent $Q/I$  profiles are also sensitive to the presence of magnetic fields inclined with respect to the symmetry axis of the incident radiation field. 

Conclusion {\bf 1} is the most reliable one because it is based on detailed 3D radiative transfer calculations of the emergent $Q/I$ for the Sr {\sc i} 4607 \AA\ line in a realistic 3D hydrodynamical model of the thermal and density structure of the quiet photosphere \citep[][]{trujillobueno-jtbnature04,trujillobueno-jtbshu07}.  
Assuming that the ``hidden'' field of the quiet solar photosphere is randomly oriented below the mean free path of the line-center photons is indeed a suitable approximation for estimating $\langle B \rangle$ \citep[e.g.,][]{trujillobueno-anusha}. Moreover, calculations based on the assumption that the magnetic field is instead horizontal also lead to the conclusion of a substantial amount of magnetic energy in the bulk of the quiet solar photosphere \citep[see \S4 in][]{trujillobueno-jtbspw4}. As reviewed in the just quoted paper several other investigations strongly support this conclusion \citep[see also the very recent contribution by][]{trujillobueno-danilovic10}.

Conclusions {\bf 2} and {\bf 3} are based on radiative transfer modeling of the $Q/I$ profile of the H$\alpha$ line observed by \cite{trujillobueno-gan00} in a quiet region at about 5$''$ from the solar limb, using various (hot and cool) 1D semi-empirical models. Any such 1D model is certainly a poor representation of the chromospheric thermal and density conditions. Fortunately,  the observed $Q/I$ profile shows a peculiar line core asymmetry which is absent in the observed $I(\lambda)$ profile. Moreover, the $Q/I$ profile of the (photo-ionization dominated) H${\alpha}$ line is not very sensitive to the chromospheric thermal structure. As shown by \cite{trujillobueno-stepan10b}, the line center asymmetry in the observed $Q/I$ profile can be explained by the Hanle effect of a magnetic field in the solar atmosphere whose height variation suggests  the presence of an abrupt and significant magnetization in the upper chromosphere of the quiet Sun and a weakly magnetized plasma directly underneath. Given that the solar chromosphere is devastatingly inhomogeneous and dynamic we cannot exclude the possibility of an alternative explanation. Nevertheless, there are other spectropolarimetric investigations that favor a sizable quiet Sun magnetization at a height of about 2000 km above the visible solar surface \citep[e.g.,][]{trujillobueno-jtbspicules,trujillobueno-holzreuter06,trujillobueno-asensiospw5,trujillobueno-centeno10} and a weakly magnetized lower chromosphere \citep[][]{trujillobueno-landi98,trujillobueno-belluzzi07}. 

Clearly, understanding the variation with height of the mean magnetization of the quiet solar chromosphere requires taking into account the multi-scale contributions of the network and internetwork magnetic loop-like structures.

\begin{acknowledgements}
The results on chromospheric magnetism reviewed here owe much to ongoing collaborations with R. Manso Sainz (IAC), R. Ramelli (IRSOL) and J. {\v{S}t\v{e}p\'an} (IAC), and I thank them for many fruitful discussions. Financial support by the Spanish Ministry of Science and Innovation through project AYA2007-63881 and by the SOLAIRE network (MTRN-CT-2006-035484) is gratefully acknowledged.
\end{acknowledgements}

\bibliographystyle{aa}

\end{document}